\documentclass[12pt]{iopart}

\usepackage{iopams}  
\usepackage{graphicx}
\usepackage{color}
\usepackage{paralist}

\begin{document}

\title[]{Employing Circuit QED to Measure Nonequilibrium Work Fluctuations}

\author{Michele Campisi $^1$, Ralf Blattmann$^1$,
Sigmund Kohler$^2$, David Zueco$^3$, and Peter H\"anggi$^1$}
\address{$^1$ Institute of Physics, University of Augsburg,
  Universit\"atsstr. 1, D-86135 Augsburg, Germany}
\address{$^2$ Instituto de Ciencia de Materiales de Madrid, CSIC, Cantoblanco, 28049 Madrid,
Spain}
\address{$^3$ Instituto de Ciencia de Materiales de Arag\'on y Departamento de F\'{i}sica de la Materia Condensada, CSIC-Universidad de Zaragoza, 50009 Zaragoza, Spain and Fundaci\'{o}n ARAID, Paseo Mar\'{i}a Agust\'{i}n 36, 50004 Zaragoza, Spain}

\ead{michele.campisi@physik.uni-augsburg.de}
\begin{abstract}
We study an interferometric method for measuring the statistics of
work performed on a driven quantum system, which has been put forward recently
[Dorner {et al.}, Phys. Rev. Lett. {\bf 110} 230601 (2013), Mazzola  {et al.},
Phys. Rev. Lett. {\bf 110} 230602 (2013)].  The method allows replacing two
projective measurements of the energy of the driven system with qubit
tomography of an ancilla that is appropriately coupled to it. We highlight
that this method could be employed to obtain the work statistics of closed as
well as open driven system, even in the strongly dissipative regime.  We then
illustrate an implementation of the method in a circuit  QED setup, which
allows one to experimentally obtain the work statistics of a parametrically driven
harmonic oscillator. Our implementation is an extension of the original method, in
which two ancilla-qubits are employed and the work statistics is retrieved
through two-qubit state tomography.  Our simulations demonstrate the
experimental feasibility.
\end{abstract}

\maketitle
\section{Introduction}
In the last two decades, the field of nonequilibrium statistical mechanics and 
thermodynamics has received a great momentum in its development due to the 
discovery of exact results, known by now as fluctuation relations. They 
characterize non-equilibrium phenomena in small systems well beyond the regime 
of linear response (in fact, to any order in the perturbative expansion) and pose 
stringent conditions on the form that the statistics of non-equilibrium fluctuating 
quantities, such as work and heat, can assume 
\cite{Bochkov77SPJETP45,Esposito09RMP81,Campisi11RMP83,Campisi11RMP83b,Jarzynski11ARCMP2,Seifert12RPP75}. 
For example, the statistics $p[w,\lambda]$ of work $w$, 
performed by varying an external parameter in a time span $[0,\tau]$ according to 
some pre specified protocol $\lambda$, is related to the statistics $p[w,\widetilde 
\lambda]$ performed when applying the time-reversed protocol $\widetilde 
\lambda$, by the formula (Tasaki-Crooks relation)
\begin{equation}
\frac{p[w,\lambda]}{p[-w,\widetilde \lambda]}= e^{\beta (w-\Delta F)}\, .
\label{eq:TC}
\end{equation}
Here the initial state of the forward (backward) process is a thermal state of  
temperature $1/k_B \beta$ and parameter value $\lambda_0$($\widetilde 
\lambda_0 = \lambda_\tau$) where $k_B$ is Boltzmann's constant, $\beta$ is the 
thermal energy and $\Delta F$ is the difference of free energy between the initial 
state of the backward protocol and the initial state of the forward protocol. We 
follow the notation of  \cite{Campisi11RMP83}, where $\lambda$, without 
subscript, denotes the function specifying the value $\lambda_t$ of the parameter 
at each time $t \in [0,\tau]$, and square brackets refer to the \emph{functional} 
dependence. Similar expressions called exchange fluctuation relations 
\cite{Campisi11RMP83,Jarzynski04PRL92,Andrieux09NJP11} hold in transport 
scenarios, where one looks at the statistics of energy and/or particles transferred 
between two reservoirs at different temperature and/or chemical potential.

Classically, the Tasaki-Crooks relation (\ref{eq:TC}) has been tested in single 
molecule stretching experiments, where they have been used to obtain the free 
energy landscape from nonequilibrium work measurements 
\cite{Liphardt02SCIENCE296,Collin05NAT437,Harris07PRL99}. In contrast, the 
experimental verification in the quantum regime is very  challenging. The problem 
lies in the fact that work is not an ordinary quantum mechanical observable 
\cite{Campisi11RMP83b,Talkner07PRE75}. It cannot be obtained by a single 
projective measurement but rather by two projective measurements of the initial 
Hamiltonian $H(\lambda_0)$ at time $t= 0$, and of the final Hamiltonian 
$H(\lambda_\tau)$ at  time $t=\tau$. The work is then given by the difference of 
the measured eigenenergies $w=E_m^{\lambda_\tau}-E_n^{\lambda_0}$. 
\footnote{Recently, new quantum fluctuation relations have been found that do not 
involve projective measurements but focus instead on the change of the quantum 
expectation of the Hamiltonian \cite{Campisi13arXiv}. This type of relation is not 
investigated here.} Huber \emph{et al.} \cite{Huber08PRL101} have proposed an 
experiment with trapped ions based on this two-measurement scheme but it 
has not been realized so far. It is worth emphasizing that such experiments would be very 
important especially because they will provide technological solutions to 
experimentally access the work statistics $p[w,\lambda]$, which is a basic building 
block for the study of thermodynamics in the quantum regime. It is central for the 
investigation of, e.g., the thermodynamic cost of quantum operations, such as 
quantum gates, which form the basis of quantum computation and quantum 
information processing \cite{Vedral02RMP74}.

One possible strategy to overcome the difficulties that the two measurements 
scheme poses has been proposed in Refs.\ 
\cite{Campisi10PRL105,Campisi11PRE83}. There the authors noted that 
intermediate quantum measurements of arbitrary observables do not alter the 
validity of the fluctuation relations, thus one might be able to retrieve the wanted 
information from  continuously monitoring some properly chosen quantum 
observable representing the flux of the wanted quantity.  As shown in 
Ref.~\cite{Campisi10PRL105} this is actually what one does in experiments of bi-
directional counting statistics \cite{Fujisawa06SCIENCE312}, where indeed the 
exchange fluctuation relation for electron transport has been verified 
experimentally by looking at the number of electrons crossing an interface 
\cite{Utsumi10PRB81,Kung12PRX2}.

Recently, Dorner \emph{et al.} \cite{Dorner13PRL110} and Mazzola \emph{et al.} 
\cite{Mazzola13PRL110} have put forward a promising method for the 
measurement of work statistics that avoids the projective energy 
measurements, and replaces them with state tomography of a qubit (the ``ancilla'') 
that is appropriately coupled to the driven system. This possibility was anticipated 
by Silva, who first pointed out the formal equivalence between the work 
characteristic function (the Fourier transform of the work statistics) and the 
Loschmidt echo \cite{Silva08PRL101}. The proposed implementations use trapped 
ions \cite{Dorner13PRL110}, and micro or nano-beams coupled to a qubit 
\cite{Mazzola13PRL110} while an experiment has just been performed using 
a nuclear magnetic resonance system \cite{Bathalao13arXiv}.

With this contribution, we (i) review the method of Dorner \emph{et al.}
\cite{Dorner13PRL110} and Mazzola \emph{et al.} \cite{Mazzola13PRL110} (Sec.\
\ref{sec:method}), (ii) discuss important extensions thereof (Sec.\
\ref{sec:extensions}), and (iii) illustrate an implementation using a circuit
QED setup (Sec.\ \ref{sec:cCQED}).  Most notably, as we shall discuss in Sec.
\ref{sec:extensions}, this new method offers a very promising tool for
accessing the work statistics of systems that are strongly coupled to their
environment \cite{Campisi09PRL102}.  This is very important because so far no
experimentally feasible method was known for this case.

The expression ``circuit QED'' refers to solid state devices that realize on a
solid-state micro-chip \cite{YouPRB03, Blais04PRA69} the physics of an atom interacting
with a light mode in a cavity, a classic problem of quantum optics
\cite{Schleich01Book}. Here the role of the atom is played by a superconducting
qubit, and the cavity is formed by a planar wave-guide. Such devices have
undergone a tremendous and fast development in the last decade
\cite{You11NATURE474,Buluta11RPP74,XiangRMP13}, allowing for the experimental study of
light-matter physics in parameter regimes that standard quantum-optics
experiments cannot reach \cite{Niemczyk10NATPHYS10,Forn-Diaz10PRL105} 
and with
an unprecedented flexibility. For example, in a circuit QED device, one can
easily manipulate the level spacing of the qubit, which can span a whole range
of values, from being resonant with the oscillator, to being far detuned from
it. This allows for manipulation of the oscillator state and its read out. For
example, following a theoretical proposal \cite{LiuEPL04},
Refs.~\cite{Hofheinz08NAT454,Hofheinz09NAT459} report on the qubit-assisted
creation and read-out of Fock states and superpositions thereof.
Importantly enough for this work, experiments have demonstrated full
two qubit tomography \cite{Filipp09PRL102b}.
Moreover, circuit QED may be used for studying thermodynamic effects on the
quantum scale \cite{QuanPRL06,QuanPRE07}.

Given the high flexibility offered by the state of the art in circuit QED, we
believe that it constitutes a very promising tool-box not only for the
development of quantum manipulation and read-out, but also for the study of its
thermodynamic cost.  The latter is an aspect which has been seldom addressed 
so
far, but is important in order to achieve quantum computers that are not only
efficient with respect to the accuracy of the logical quantum gates, but also
with respect to avoiding detrimental heating.  Here
we suggest a circuit QED implementation for the measurement of work statistics,
which constitutes a first step towards the development of this tool-box. 

\section{The method}
\label{sec:method}
In this section we shall briefly review the method for extracting the work statistics
put forward by Dorner \emph{et al.} \cite{Dorner13PRL110} 
and Mazzola \emph{et al.} \cite{Mazzola13PRL110}. We shall 
follow primarily the presentation given in Ref.~\cite{Dorner13PRL110}.

Given a driven quantum system described by the Hamiltonian
\begin{equation}
H_S(\lambda_t) = H_0 - \lambda_t Q ,
\label{eq:H-S}
\end{equation}
the aim of the method is to provide an experimentally feasible prescription of how  
one can obtain its work statistics in an experiment. Here $\lambda_t$ and $Q$ 
denote an externally applied generalized force and its  conjugate displacement, 
respectively. $Q$ is a quantum mechanical observable, whereas $ \lambda_t$ is 
a classical quantity, whose evolution in time from $t=0$ to $t=\tau$ is pre-specified  
\cite{Kubo57aJPSJ12}.  Prototypical examples are a forced oscillator 
\cite{Campisi08PRE78b,Talkner08PRE78}, and a parametrically driven oscillator, 
i.e., an oscillator with a time dependent frequency 
\cite{Deffner08PRE77,Husimi53PTP9}.

The traditional prescription requires that the system is prepared at time $t=0$ in 
the thermal state:
\begin{equation}
\rho_S= e^{- \beta H(\lambda_0)}/Z_S(\lambda_0) ,
\end{equation}
where $\beta$ is the inverse thermal energy and $Z_S(\lambda_0)=\Tr  e^{-\beta
H(\lambda_0)} $.  Projective measurements of $H(\lambda_0)$ and
$H(\lambda_\tau)$ are then performed at times $t=0$ and $t=\tau$, providing
us with one eigenvalue of the initial Hamiltonian and one of the final
Hamiltonian, $E_n^{\lambda_0}$ and $E_m^{\lambda_\tau}$, respectively.  The 
work $w=E_m^{\lambda_\tau}-E_n^{\lambda_0}$ is then recorded, so that repeated
measurements allow one to sample the work probability distribution function
\begin{equation}
 p[w;\lambda]= \sum_{m,n}
\delta(w-(E_m^{\lambda_\tau}-E_n^{\lambda_0}))p_{m|n}[\lambda]e^{-\beta 
E_n^{\lambda_0}}/Z_0 ,
 \label{eq:p[w;lambda]}
 \end{equation}
where $p_{m|n}[\lambda]$ denotes the transition probability from state
$n$ to state $m$ induced by the protocol $\lambda$.

The major obstacle for implementing this prescription comes from the
experimental difficulty to perform projective measurements on the system of
interest. The method of Dorner \emph{et al.} \cite{Dorner13PRL110} and
Mazzola \emph{et al.} \cite{Mazzola13PRL110} circumvents this difficulty
by coupling the system to an ``ancilla'', namely a qubit, which is used to
read out the Fourier transform $G[u,\lambda]$ of the work statistics
$p[w,\lambda]$ \cite{Talkner07PRE75}:
\begin{eqnarray}
G[u,\lambda] = \int dw e^{iuw}p[w;\lambda] &=&
\langle U_S^\dagger[\lambda]  e^{iu H(\lambda_\tau)/\hbar} U_S[\lambda] e^{-iu 
H(\lambda_0)/\hbar} \rangle_S \nonumber \\
&=& \langle (e^{-iu H(\lambda_\tau)/\hbar} U_S[\lambda])^\dagger U_S[\lambda] 
e^{-iu H(\lambda_0)/\hbar} \rangle_S .
\label{eq:G[u;lambda]}
\end{eqnarray}
Here $U_S[\lambda]$ is the time evolution operator generated by the driving
protocol $\lambda$, and  $\langle \cdot \rangle_S$ denotes average over
$\rho_S$. Note that since $P[w,\lambda]$ is a real function, the relation
$G[-u,\lambda]=G^*[u,\lambda]$ holds.

Following  Dorner \emph{et al.} \cite{Dorner13PRL110}, the system (S) is coupled 
to the ancilla (A) according to the Hamiltonian
\begin{equation}
H_{S+A} = \frac{\hbar\varepsilon}{2}\sigma^z + H_0 - (\chi^+_t \Pi_+  +
\chi^-_t \Pi_-) Q ,
\label{eq:H_A+S}
\end{equation}
where $\sigma^z=\Pi_+ -\Pi_-$ is the $z$-Pauli matrix, $\Pi_{\pm}=
|\pm\rangle \langle \pm|$ is the projector onto the ancilla states
$|\pm\rangle$,  and $\chi^\pm_t$ are two independent driving protocols of
duration $T=\tau+u$ which will be specified later.

Because the system-ancilla coupling commutes with the free ancilla
Hamiltonian $H_A= \varepsilon\sigma^z/2$, the evolution
$U_{S+A}[\chi^+,\chi^-]$ of $S+A$ generated by the drivings $\chi^+,\chi^-$
is block diagonal in the basis $\{|+\rangle, |-\rangle \}$:
\begin{equation}
U_{S+A}[\chi^+,\chi^-] = \left(\begin{array}{cc}e^{-i\varepsilon T/2\hbar
}U_S[\chi^+] & 0 \\0 & e^{i\varepsilon T/2\hbar
}U_S[\chi^-]\end{array}\right) .
\label{eq:blockU}
\end{equation}
Choosing
\begin{equation}
\chi^+_t=\left\{ \begin{array}{lll}
\lambda_t \quad $for $ t \in [0,\tau] \\
\lambda_\tau \quad $for $t \in [\tau,\tau+u]
\end{array}\right. , \quad
\chi^-_t=\left\{ \begin{array}{lll}
\lambda_0 \quad $for $ t \in [0,u] \\
\lambda_{t-u} \quad $for $t \in [u,\tau+u]
\end{array}\right. ,
\label{eq:g_pm}
\end{equation}
the time evolution operator reads
\begin{equation}
U_{S+A}[\chi^+,\chi^-] = \left(\begin{array}{cc}e^{-i\varepsilon (\tau+u)/2\hbar } e^{-
iu H(\lambda_\tau)/\hbar} U_S[\lambda] & 0 \\
0 & e^{i\varepsilon (\tau+u)/2\hbar }U_S[\lambda] e^{-iu
H(\lambda_0)/\hbar}\end{array}\right) .
\end{equation}
It contains the operators $U_S[\lambda] e^{-iu H(\lambda_0)/\hbar}$ and
$e^{-iu H(\lambda_\tau)/\hbar} U_S[\lambda]$ which appear in the expression
of the characteristic function, Eq. (\ref{eq:G[u;lambda]}).

Inspired by Ramsey interferometry, the idea is to prepare the system in a
superposition of up and down states so that the two time evolutions interfere
and the wanted information will be encoded in the state of the ancilla at
the final time $T=\tau +u$.  This is achieved by the following
protocol \cite{Dorner13PRL110,Mazzola13PRL110}:
\begin{compactenum}[1.]
\item Prepare the compound system in the state $\rho_{S+A}= |-\rangle \langle -| 
\rho_S$ at $t<0$.
\item Perform a Hadamard operation $\sigma^H=(\sigma^x+\sigma^z)/\sqrt{2}$ on the 
ancilla at $t=0$.
\item Let the $S+A$ system evolve for a time $\tau+u$ according to $U_{S+A}[\chi^+,
\chi^-]$.
\item Perform a Hadamard operation $\sigma^H$ on the ancilla at $t=T=\tau+u$.
\end{compactenum}

After this protocol, the ancilla is described by the reduced density matrix
\begin{eqnarray}
\rho_A(u) &=& \Tr_S \sigma^H U_{S+A} [\chi^+,\chi^-]\sigma^H \rho_{S+A} 
\sigma^H U^\dagger_{S+A}[\chi^+,\chi^-]  \sigma^H
\\ &=&  \left(1- [\Re L(u)] \sigma^z + [\Im L(u)]\sigma^y \right)/2 ,
\end{eqnarray}
where $\Re$ and $\Im$ denote real and imaginary parts, $\Tr_S$ is the trace over 
the system Hilbert space, and
\begin{eqnarray}
L(u)= e^{-i\varepsilon (\tau + u)/\hbar } G[u,\lambda] .
\end{eqnarray}
Thus, by state tomography of the ancilla density matrix at time $t=T=\tau+u$ one 
can recover the value of the characteristic function $G[u,\lambda]$ for a given $u$. By 
repeating the whole procedure for various values of $u \in (0, \infty)$, one obtains 
$G[u,\lambda]$ on the positive 
real axis. Using $G[-u,\lambda]=G^*[u,\lambda]$ one obtains  $G[u,\lambda]$ on 
the whole real axis, and then, by inverse Fourier transform the work statistics $p[w,
\lambda]$. In practice one can sample the characteristic function only on a finite domain.
This, in turn, limits the accuracy with which the work probability distribution
function can be resolved.

The purpose of the first Hadamard transformation is to create a
superposition of the up and down states.  The second Hadamard
recombines the entries of the ancilla density matrix at time $T=\tau+u$
and, hence it is not strictly necessary.

It is worth emphasizing that the diagonal coupling in Eq.~(\ref{eq:H_A+S}) can
often be realized only approximately. The implementation proposed in
Sec.~\ref{sec:cCQED} is one example of an approximate diagonal
coupling obtained by far detuning the ancilla with respect to the system's
transition energies. 
With the perfectly diagonal coupling, the wavefunction is prepared in a superposition
in which one component follows the forward protocol $\chi^+$, while the other follows
the backward protocol $\chi^-$.  While both components evolve independent of each
other with the respective $U_S[\chi^\pm]$ of the isolated system, the ancilla collects the
resulting phase difference.

\section{Important extensions}
\label{sec:extensions}

\subsection{Work statistics of arbitrary open quantum systems}
As mentioned above the primary advantage of the interferometric scheme of 
Dorner \emph{et al.} \cite{Dorner13PRL110} and Mazzola  \emph{et al.} 
\cite{Mazzola13PRL110} is that it avoids projective measurements on the system 
of interest $H_S$ by replacing them with
state tomography on the ancilla. This has a very important consequence in regard 
to the possibility of experimentally testing fluctuation theorems in open quantum 
systems with arbitrarily strong coupling to a thermal environment 
\cite{Campisi09PRL102}
\begin{equation}
H_{S+B}(\lambda_t) = H_S(\lambda_t) + H_B + H_{SB} \, .
\end{equation}
Here $H_B$ is the thermal bath Hamiltonian and $H_{SB}$ is an arbitrarily strong 
coupling.
Nevertheless, the fluctuation theorem continues to hold unaltered in this
case \cite{Campisi09PRL102}, because when driving the system
$S$, part of the injected energy may flow to the bath $B$ and in the $SB$
interaction. Thus the work spent to drive the system is given by the change
in the $S+B$ energy: $w= E_{S+B,m}^{\lambda_\tau}- E_{S+B,n}^{\lambda_0} $.
But since one can see the $S+B$ system as a closed system staying initially
in a thermal state with (common) inverse temperature $\beta$,
\begin{equation}
\rho_{S+B} = \frac{e^{-\beta H_{S+B}(\lambda_0)}}{Z_{S+B}(\lambda_0)}\, ,
 \end{equation}
the ordinary fluctuation relation applies $P[w,\lambda]/P[-w,\widetilde{\lambda}]= 
e^{\beta (w-\Delta F_{S+B})}$
independent of the coupling strength.
Using the expression of the free energy of an arbitrary open quantum system 
\cite{Feynman63AP24}
$F_S(\lambda_t) = F_{S+B}(\lambda_t)-F_B^0$ (where $F_B^0=-\beta^{-1} \ln 
\Tr_B e^{-\beta H_B}$, and $F_{S+B}(\lambda_t)=-\beta^{-1} \ln \Tr_{S+B} e^{-
\beta H_{S+B}(\lambda_t)}$), one immediately sees that $\Delta F_{S+B}=F_{S
+B}(\lambda_t)-F_{S+B}(\lambda_0)
= F_{S}(\lambda_t)-F_{S}(\lambda_0)= \Delta F_{S}$. Thus the fluctuation 
theorem remains unaltered in the case of arbitrary open quantum systems 
\cite{Campisi09PRL102}:
\begin{equation}
\frac{P[w,\lambda]}{P[-w,\widetilde{\lambda}]}= e^{\beta(w-\Delta F_{S})}\, .
 \end{equation}
This result is the quantum version of a result obtained by
Jarzynski for classical systems \cite{Jarzynski04JSM04}.  The main
difference between the classical and the quantum case is that while in the
classical case one may obtain the work $w$ performed on the $S+B$ system by
looking at the trajectory of $S$ alone  \cite{Jarzynski04JSM04}, in the
quantum case this is impossible \cite{Campisi11RMP83}. In the quantum
case, in principle, one should perform two projective measurements of the
total Hamiltonian $H_{S+B}$. Making a projective measurement on $S$ alone
is already a challenging task in many experimental setups; making a
projective measurement of $S+B$ seems much more difficult, if not
impossible. The interferometric scheme may be effective in overcoming this
problem. If now the open system is coupled to the ancilla which, in turn,
has no direct contact to the environment, the $S+B+A$ Hamiltonian reads
\begin{equation}
H_{S+B+A} = \frac{\hbar\varepsilon}{2}\sigma^z + H_0 + H_B + H_{SB} -
(\chi^+_t \Pi_+  + \chi^-_t \Pi_-) Q .
\label{eq:H_S+B+A}
\end{equation}
Implementing the same interferometric scheme as in Sec.~\ref{sec:method}
with the initial state $\Pi_- \rho_{S+B}$ results in the
characteristic function of the open system $H_{S+B}(\lambda_t)$. Thus the
interferometric approach provides, if the ancilla is well isolated, access
to the work distribution of arbitrary open as well as closed nonequilibrium
quantum systems.

Most remarkably, our present discussion highlights that in the
interferometric scheme of Dorner \emph{et al.} \cite{Dorner13PRL110} and
Mazzola  \emph{et al.} \cite{Mazzola13PRL110}, deviations from the
fluctuation theorem are expected only as a consequence of thermal noise on
the ancilla $A$. Thermal noise on the system $S$ may affect the statistics
of work itself, but not the validity  of the fluctuation relations.

We emphasize that the fluctuation theorem for open quantum systems
described in this section is fully general and exact. In particular it does
not require the interaction $H_{SB}$ to be weak nor the initial $S+B$ state
to be uncorrelated. Quite on the contrary, in case of strong coupling the
initial state $\rho_{S+B}$ contains correlations, and the subsequent
evolution of the reduced system density matrix needs not be described by
completely positive maps \cite{Pechukas94PRL73}, nor has to be Markovian.
In this regard newly introduced definitions of characteristic functions for
open quantum systems in terms of the reduced system dynamics
\cite{Kafri12PRA86,Rastegin13arxiv,Albash12arXiv,Chetrite12JSP148} must be
regarded as approximate expressions whose validity is not
guaranteed, and whose main object is generally not the work (i.e., the
change in energy of $S+B$) but some other quantity that pertains to the
system $S$ only.

\subsection{Exclusive vs.\ inclusive work statistics}
Fluctuation relations appear in the literature in two
complementary ways, referred to as exclusive and inclusive viewpoint
\cite{Jarzynski07CRPHYS8,Campisi11PTRSA369}. In our discussion so far we
have been adopting the inclusive viewpoint, which addresses the probability
of the change $w$ in the system energy $H_0-\lambda_t Q$, i.e., including
the driving term $\lambda_t Q$.  One may want to look also at the
statistics $p_0[w_0,\lambda]$ of the change $w_0$  in the energy of the system
$H_0$ excluding the driving energy $\lambda_t Q$ \footnote{For the sake of
clarity we stress that the dynamics $U[\lambda]$ is one and the same in
both viewpoints. What changes is only what one looks at, namely $w$ in the
inclusive case and $w_0$ in the exclusive case.}. This is given by $w_0=
e_m-e_n$, where $e_{m}$ and $e_n$ are eigenvalues of $H_0$ obtained by
projective measurements of $H_0$ at $t=0$ and at $t=\tau$, respectively
\cite{Campisi11PTRSA369}.  The exclusive fluctuation relation reads
\cite{Bochkov77SPJETP45,Campisi11PTRSA369}
\begin{equation}
\frac{p_0[w_0,\lambda]}{p_0[-w_0,\widetilde{\lambda}]}= e^{\beta w_0},
\end{equation}
while the characteristic function of exclusive work is given by
\cite{Campisi11PTRSA369}
\begin{eqnarray}
G_0[u,\lambda] = \int dw e^{iuw}p_0[w;\lambda] &=&
\langle U_S^\dagger[\lambda]  e^{iu H_0/\hbar} U_S[\lambda] e^{-iu H_0/\hbar} 
\rangle_S \nonumber \\
&=& \langle (e^{-iu H_0/\hbar} U_S[\lambda])^\dagger U_S[\lambda] e^{-iu 
H_0/\hbar} \rangle_S .
\label{eq:G[u;lambda]-excl}
\end{eqnarray}
It differs from the exclusive work characteristic function in
Eq.~(\ref{eq:G[u;lambda]}) in that $e^{-iu H_0/\hbar}$ appears instead of
$e^{-iu H(\lambda_\tau)/\hbar}$.  Accordingly, $G_0[u,\lambda]$ can be
accessed by means of the interferometric scheme described in Sec.
\ref{sec:method} by replacing in Eq.~(\ref{eq:g_pm}) the drivings $\chi^\pm$ by
\begin{equation}
\chi^+_t=\left\{ \begin{array}{lll}
\lambda_t \quad $for $ t \in [0,\tau] \\
0 \quad $for $t \in [\tau,\tau+u]
\end{array}\right. , \quad
\chi^-_t=\left\{ \begin{array}{lll}
0 \quad $for $ t \in [0,u] \\
\lambda_{t-u} \quad $for $t \in [u,\tau+u]
\end{array}\right. ,
\label{eq:g_pm_excl}
\end{equation} 
so that the evolution takes on the form
\begin{equation}
U_{S+A}[\chi^+,\chi^-] = \left(\begin{array}{cc}e^{-i\varepsilon (\tau+u)/2\hbar } e^{-
iu H_0/\hbar} U_S[\lambda] & 0 \\
0 & e^{i\varepsilon (\tau+u)/2\hbar }U_S[\lambda] e^{-iu H_0/\hbar}\end{array}
\right) ,
\end{equation}
where we recognize the operators $U_S[\lambda] e^{-iu H_0/\hbar}$ and $e^{-iu 
H_0/\hbar} U_S[\lambda]$,
appearing in the expression of the exclusive characteristic function, Eq. 
(\ref{eq:G[u;lambda]-excl}). Accordingly the state of the ancilla
at time $\tau+u$ encodes the information on the exclusive 
characteristic function $G_0[u,\lambda]$.

\section{Circuit QED implementation}
\label{sec:cCQED}
We want to experimentally access the work statistics of a parametrically
driven quantum oscillator, whose frequency changes in time according to
$\omega^2(t)= \omega^2- 4 \omega \lambda_t $. Its Hamiltonian reads
\begin{eqnarray}
H_S(\lambda_t) &=& p^2/2m + m(\omega^2- 4 \omega \lambda_t) x^2/2 
\nonumber \\
&=& \hbar\omega (a^\dagger a +1/2)- \hbar \lambda_t (a^\dagger +a)^2 ,
\label{eq:process}
\end{eqnarray}
where $a= x \sqrt{m \omega/2\hbar} + i p \sqrt{1 /2m \omega \hbar}$ and
$a^\dagger= x \sqrt{m \omega/2\hbar} - i p \sqrt{1 /2m \omega \hbar}$ are
the usual bosonic shift operators.
To implement this Hamiltonian we consider a circuit QED setup, 
where a qubit is coupled to a single mode $\omega$ of a line resonator 
\cite{Blais04PRA69}.
The qubit-oscillator system is described by the Rabi Hamiltonian 
\begin{eqnarray}
H_{S+A} &= \hbar \varepsilon \sigma^z/2 + \hbar \omega(a^\dagger a+1/2)+
\hbar g(a^\dagger +a)\sigma^x ,
\label{eq:rabi}
\end{eqnarray}
where $\sigma^x$, $\sigma^z$ are Pauli matrices, $\varepsilon$ is the qubit
energy splitting, and $g$ is the qubit-oscillator interaction strength.
Note that this Hamiltonian is generally not of the type of that in
Eq.~(\ref{eq:H_A+S}). First, the  qubit-system interaction does not commute
with the free qubit Hamiltonian $\hbar \varepsilon \sigma^z/2$. Second, the
interaction term is linear in $a^\dagger +a$, whereas we aim at implementing
an interaction quadratic in $a^\dagger +a$. Third, typical circuit QED
setups do not provide the possibility to control the interaction $g$ in
time, because  $g$ is fixed by the geometry of the device. With current
technology \cite{Blais04PRA69} one can relatively easily control the qubit
splitting $\varepsilon$, while setups allowing for the control of $g$ so far have
been studied only theoretically \cite{Peropadre10PRL105,Reuther11NJP9}. 
A full description of the transmission line would contain also higher harmonics
at multiples of the fundamental frequency $\omega$.  The protocol considered
below, however, acts directly only upon the fundamental mode, while its
harmonics are affected only indirectly via the qubit.  We therefore assume that
they remain close to their ground state and, thus, do not take them into
account.

\begin{figure}[]
	\begin{center}
		\includegraphics[width=.6\textwidth]{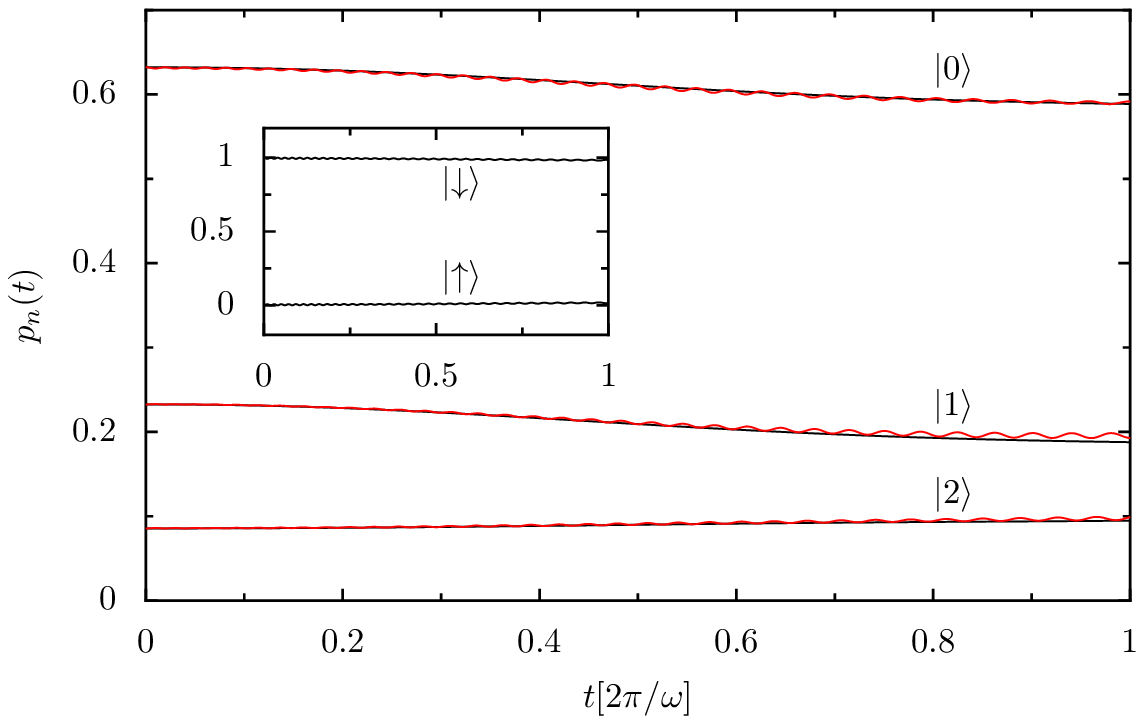}
		\caption{Comparison between the dynamics generated by the Rabi 
Hamiltonian in Eq.~(\ref{eq:rabi}), black line, and the 
				dynamics generated by the diagonal Hamiltonian in 
Eq.~(\ref{eq:rabi-g<<e-ad}), red line. The plot shows the evolution 			
	of the population of the first three eigenstates of the oscillator.
				The inset shows the corresponding evolution of the qubit 
population. The initial state was 
				$\left | \downarrow \right> \left < \downarrow \right|   e^{-\beta 
H_{S}(\lambda_0)}/\Tr e^{-\beta H_{S}(\lambda_0)} $.
				We used the driving $\varepsilon_t= \hbar g^2 / 2 \lambda_t$, 
where $\lambda_t = \lambda_0+v t$,
				and the following parameters: $g=2.5 \hbar \omega$,  
				$\lambda_0=0.0625 \hbar \omega$, $v=1.5 \lambda_0
 \omega /2\pi $, $1/\beta= \hbar \omega
$. }
		\label{fig:Fig1}
	\end{center}
\end{figure}

These issues can be partially solved by considering a time dependent qubit
splitting $\varepsilon_t$ and working in a regime where the coupling $g$
and the oscillator frequency $\omega$ are small:
\begin{equation}
g \simeq \omega \ll \varepsilon_t \,.
\end{equation}
By applying the time-dependent unitary transformation
\begin{equation}
\Omega_t= e^{i g(a^\dagger +a)\sigma_y/\varepsilon_t}
\label{eq:transf}
\end{equation}
and neglecting terms beyond second order in the small parameter $g/
\varepsilon_t$, we obtain, up to a global energy shift the Hamiltonian
\begin{eqnarray}
\hspace{-2.5cm}H_{S+A}'(\varepsilon_t) &=& \Omega_t H_{S+A} \Omega^
\dagger_t + i\dot \Omega_t \Omega_t^\dagger \label{eq:rabi-g<<e}  \\
\hspace{-2.5cm}&=& \frac{\hbar\varepsilon_t}{2} \sigma^z + \hbar \omega \left(a^
\dagger a+\frac{1}{2}\right) + \frac{\hbar g^2}{\varepsilon_t} (a^\dagger +a)^2 
\sigma^z
+ i \frac{\hbar \omega g}{\varepsilon_t}(a^\dagger-a)\sigma_y
-  \frac{ \hbar g  \dot \varepsilon_t}{\varepsilon^2_t}(a^\dagger+a)\sigma_y .
\nonumber
\end{eqnarray}
The last term comes from the explicit time dependence of the transformation $
\Omega_t$. We shall consider a qubit driving that is slow compared to the qubit's 
own time scale while being comparable to the oscillator's time scale
\begin{equation}
\dot \varepsilon_t / \varepsilon_t \simeq \omega \ll \varepsilon_t \,.
\end{equation}
In this way the oscillator can be driven out of equilibrium while the qubit undergoes 
an  adiabatic evolution.
Note that the factors $g\omega/\varepsilon_t$  and $g  \dot \varepsilon_t/
\varepsilon^2_t$ are comparable to the factor $g^2/\varepsilon_t$
appearing in the third term of Eq. (\ref{eq:rabi-g<<e}). However the last two terms 
are oscillating much faster and can therefore be neglected within a rotating-wave 
approximation. This can be seen by going to the interaction picture with respect to 
$\hbar \varepsilon_t \sigma^z /2 + \hbar (\omega a^\dagger a +1/2)$, where the 
last two terms contain the frequencies $\pm (\bar \varepsilon_t \pm \omega)\simeq 
\pm \bar \varepsilon_t=\pm t^{-1} \int_0^t \varepsilon_sds$ and  the second term 
contains the much lower frequencies $0,\pm 2\omega$. We thus conclude that 
\begin{eqnarray}
H'_{S+A} (\varepsilon_t) &=& \frac{\hbar \varepsilon_t}{2} \sigma^z + \hbar \omega 
\left(a^\dagger a +\frac{1}{2}\right)+ \frac{\hbar g^2}{\varepsilon_t} (a^\dagger 
+a)^2 \sigma^z \nonumber \\
&=& \frac{\hbar \varepsilon_t}{2} \sigma^z +\hbar \omega \left(a^\dagger a +
\frac{1}{2}\right) + 
\left(\frac{\hbar g^2}{\varepsilon_t}\Pi_+ - \frac{\hbar g^2}{\varepsilon_t} \Pi_- \right)
(a^\dagger +a)^2 
\label{eq:rabi-g<<e-ad}
\end{eqnarray}
is a good approximation of $H_{S+A}$ in the chosen parameter regime.
The form (\ref{eq:rabi-g<<e-ad}) is already rather close to the desired
Hamiltonian in Eq.~(\ref{eq:H_A+S}). The main difference is that
in Eq.~(\ref{eq:H_A+S}), one drives the two subspaces spanned by $\Pi_\pm$ with 
two independent drivings
$\chi^\pm_t$, whereas here we have only one driving parameter $\varepsilon_t$ 
that drives both subspaces at the same time.
The other difference is that now the free qubit Hamiltonian is time dependent. This 
affects only an overall phase, which 
therefore is not our major concern here. 
Figure \ref{fig:Fig1} illustrates how the original Rabi Hamiltonian (\ref{eq:rabi}) is 
well approximated by the diagonal Hamiltonian in Eq. (\ref{eq:rabi-g<<e-ad}). In 
the appendix we provide an alternative derivation of $H'_{S+A}$ based on the 
explicit calculation of the time-evolution generated by $H_{S+A}$.

Note that the transformation (\ref{eq:transf}) is similar but not quite the
same as the transformation commonly employed in the dispersive regime
\cite{Zueco09PRA80}. We might call the regime investigated here, where the
oscillator is very slow, the \emph{soft mode} regime, and the resulting
effective Hamiltonian $H'_{S+A}$, Eq.~(\ref{eq:rabi-g<<e-ad}), the
\emph{soft mode Hamiltonian}. Like the dispersive Hamiltonian, the soft
mode Hamiltonian is diagonal in the natural qubit-oscillator basis. But
while the dispersive Hamiltonian represents a qubit-oscillator
coupling linear in $(a+a^\dagger)$, the soft mode Hamiltonian describes a
coupling quadratic in $(a+a^\dagger)$.

\subsection{Introducing a second qubit}
\begin{figure}[]
	\begin{center}
		\includegraphics[width=.6\textwidth]{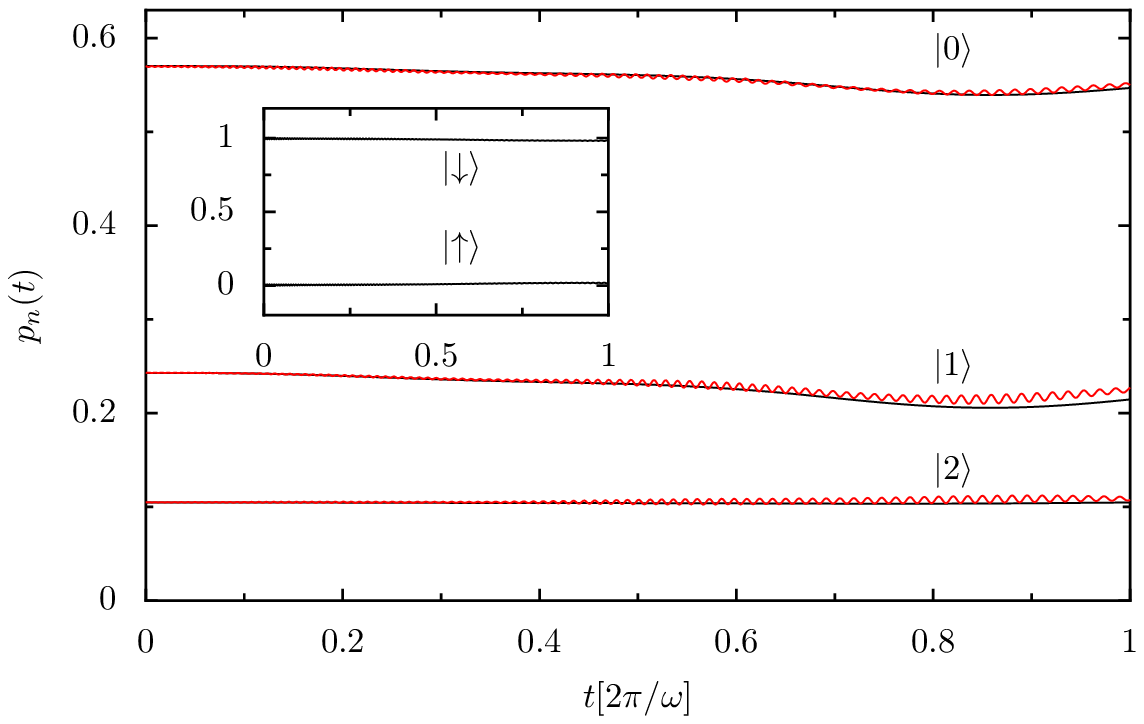}
		\caption{Comparison between the dynamics generated by the
Tavis-Cummings Hamiltonian in Eq.~(\ref{eq:rabi2}), black line, and 
				the dynamics generated by the diagonal
Hamiltonian in Eq.~(\ref{eq:rabi2-ad}), red line. The plot shows the
				evolution of the population of the first three eigenstates of the 
oscillator.
				The inset shows the corresponding evolution of the first qubit 
population. The initial state was 
				$\rho_{S+2A}$ (see Eq.~\ref{eq:rhoS2A}), and $\varepsilon_{1,t}$, 
$\varepsilon_{2,t}$ were chosen as in Fig.~\ref{fig:Fig3},
				bottom right panel, as to realize the drivings $\chi^{\pm}_t$ 
shown in Fig.~\ref{fig:Fig3},
				bottom left panel, corresponding to a linear ramp $
\lambda_t=\lambda_0+vt$.
				We used the following parameters: $g_1=2.5 \hbar 
\omega$, $g_2=0.5 \hbar \omega$, 
				$\lambda_0=0.0625 \hbar \omega$,  $v=1.5 \lambda_0 \omega / 2\pi $, $1/\beta= \hbar\omega
$.}
		\label{fig:Fig2}
	\end{center}
\end{figure}

In order to allow for the independent driving of two subspaces, we modify the 
method described above by introducing a second qubit.
The work characteristic function measurement is thus assisted by two ancillae. 
Two-qubit state tomography has been reported recently in \cite{Filipp09PRL102b}. 
Our starting Hamiltonian is:
 \begin{eqnarray}
\hspace{-1cm}H_{S+2A} = \hbar \varepsilon_1 \sigma^z_1/2 + \hbar \varepsilon_2
\sigma^z_2/2 + \hbar \omega(a^\dagger a+1/2)+ \hbar(a^\dagger
+a)(g_1\sigma^x_1+g_2\sigma^x_2) .
\label{eq:rabi2}
\end{eqnarray}
Following the derivation illustrated above, we shall work in the regime
\begin{equation}
\omega,  g_i \ll \varepsilon_{i,t}\, , \quad \dot \varepsilon_{i,t} /
\varepsilon_{i,t} \simeq \omega \ll \varepsilon_{i,t}\, , \quad i=1,2 .
\label{eq:regime2}
\end{equation}
By applying the transformation
\begin{equation}
\Omega_t = e^{ig_1(a^\dagger +a)\sigma^y_1/\varepsilon_{1,t}}e^{ig_2(a^\dagger 
+a)\sigma^y_2/\varepsilon_{2,t}}
\end{equation} 
and neglecting cubic or higher terms in $g_1/\varepsilon_1, g_2/\varepsilon_2$, as 
well as fast oscillating contributions we arrive at
\begin{eqnarray}
\hspace{-2.7cm}H'_{S+2A}(\varepsilon_{1,t},\varepsilon_{2,t}) = \frac{\hbar}
{2}\varepsilon_{1,t} \sigma^z_1 +  \frac{\hbar}{2} \varepsilon_{2,t} \sigma^z_2 + 
\hbar \omega \left(a^\dagger a + \frac{1}{2}\right)+
\hbar (a^\dagger +a)^2\left( \frac{g_1^2}{\varepsilon_{1,t} } \sigma^z_1+
\frac{g_2^2}{\varepsilon_{2,t}} \sigma^z_2\right) .
\label{eq:rabi2-ad}
\end{eqnarray}
Figure \ref{fig:Fig2} illustrates how the original Rabi Hamiltonian (\ref{eq:rabi2}) is 
well approximated by the diagonal Hamiltonian in Eq.~(\ref{eq:rabi2-ad}).
It is worthwhile rewriting $H'_{S+2A}$ in terms of projectors $\Pi_{\pm,\pm}$ onto 
the four states $|\pm, \pm \rangle$:
\begin{eqnarray}
H'_{S+2A}(\varepsilon_{1,t},\varepsilon_{2,t}) 
&=&\hbar [\varepsilon^+_t + \omega (a^\dagger a+1/2) + \chi^+_t (a^\dagger 
+a)^2]\Pi_{++} \nonumber \\
&+& \hbar [\varepsilon^-_t + \omega (a^\dagger a+1/2) + \chi^-_t (a^\dagger 
+a)^2]\Pi_{+-} \nonumber \\
&+&\hbar [-\varepsilon^-_t + \omega (a^\dagger a+1/2) - \chi^-_t (a^\dagger 
+a)^2]\Pi_{-+} \nonumber \\
&+&\hbar [-\varepsilon^+_t + \omega (a^\dagger a+1/2) - \chi^+_t (a^\dagger 
+a)^2]\Pi_{--} ,
\end{eqnarray}
where
\begin{eqnarray}
\varepsilon^{\pm}_t = \frac{\varepsilon_{1,t} \pm \varepsilon_{2,t}}{2} ,\\
\chi^{\pm}_t = \frac{g_1^2}{\varepsilon_{1,t}} \pm \frac{g_2^2}{\varepsilon_{2,t}}.
\label{eq:chip-chim}
\end{eqnarray}
By focussing onto the subspace spanned by $\Pi_{-+}$ and $\Pi_{--}$ we see that 
by manipulating the
two splittings $\varepsilon_{1,t}$ and $\varepsilon_{2,t}$, one can realize two 
independent drivings $\chi^+_t$ and
$\chi^-_t$ acting in  the respective sub-subspace.
This realizes all the ingredients that we need for implementing the characteristic 
function measurements protocol employing a circuit 
QED setup.

\subsection{The protocol}
\begin{figure}[]
	\begin{center}
		\includegraphics[width=\textwidth]{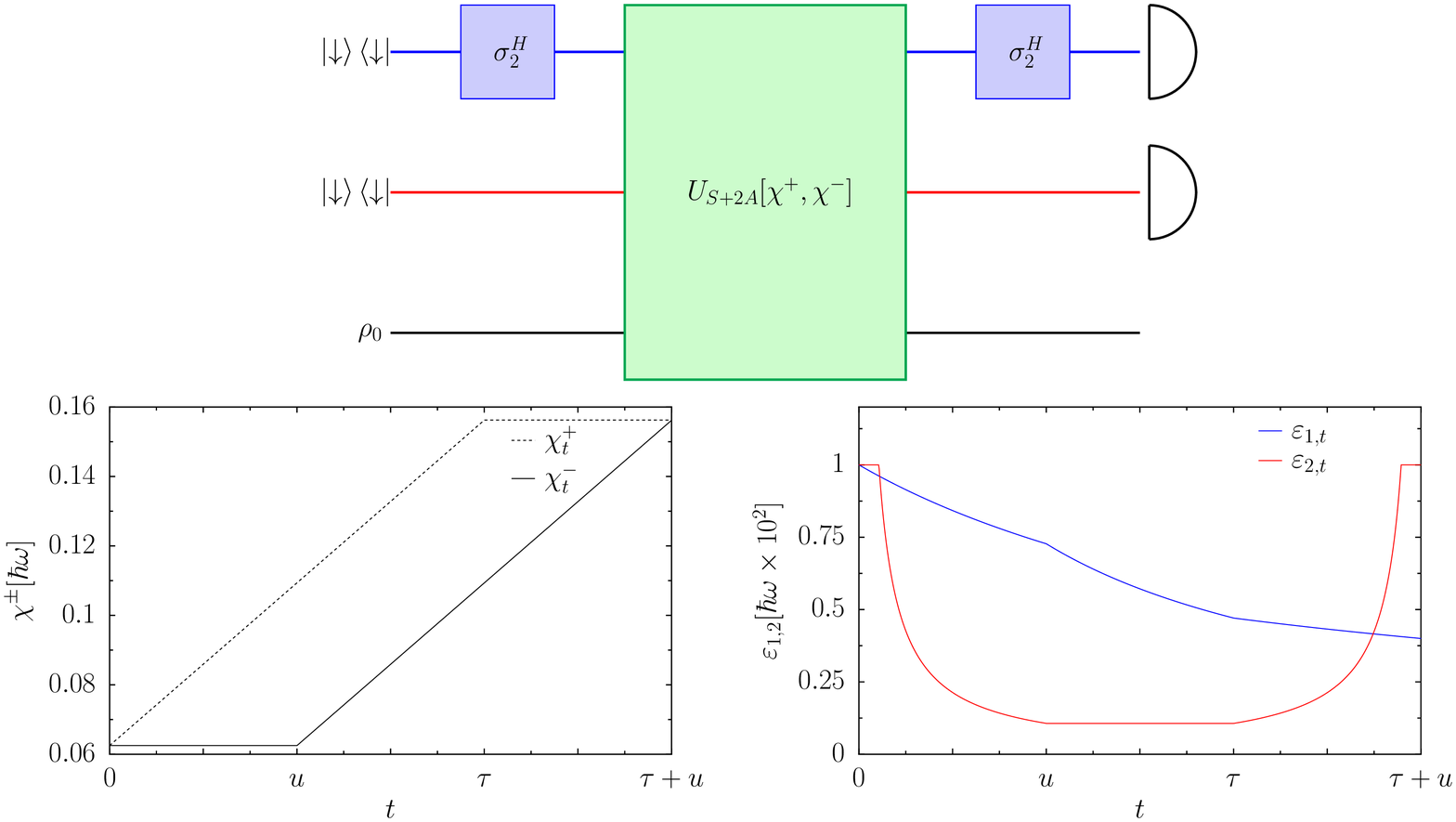}
		\caption{Top: Schematics of the two-qubit protocol. Bottom left: time 
evolution
		of the two driving parameters $\chi^{\pm}_t$, Eq. (\ref{eq:g_pm}), for a 
linear ramp of $\lambda_t$. Bottom right: the time evolution
		of the two qubit splittings $\varepsilon_{i,t}$ that realize the drivings  $
\chi^{\pm}_t$, see Eq. (\ref{eq:vareps12}).}
		\label{fig:Fig3}
	\end{center}
\end{figure}

First, the two drivings $\varepsilon_{1,t}$ and $\varepsilon_{2,t}$ are chosen in
such a way as to realize the protocols $\chi^+_t,\chi^-_t$ in Eq. (\ref{eq:g_pm}). 
This is achieved by solving Eq. (\ref{eq:chip-chim}) for $\varepsilon_{1,t},
\varepsilon_{2,t}$ to obtain:
\begin{equation}
\varepsilon_{1,t} = \frac{2g_1^2}{ \chi^{+}_t+ \chi^{-}_t} \, , \quad \varepsilon_{2,t} = 
\frac{2g_2^2}{ \chi^{+}_t- \chi^{-}_t} .
\label{eq:vareps12}
\end{equation}
With this choice, the protocol goes as follows (see Fig. \ref{fig:Fig3} top
panel):

\begin{compactenum}[1.]
\item Prepare the system at $t<0$ in the state:
\begin{equation}
\rho_{S+2A} = \frac{e^{-\beta (\omega a^\dagger a - \lambda_0(a^\dagger+a)^2)}}
{Z_{S+2A}(\lambda_0)} \Pi_{--} .
\label{eq:rhoS2A}
\end{equation}
\item Perform a Hadamard operation $\sigma^H_2= (\sigma^x_2+\sigma^z_2)/\sqrt{2} 
$ on the second qubit at time $t=0$.
\item Let the $S+2A$ system evolve for a time $\tau+u$ according to $H_{S+2A}
(\varepsilon_{1,t},\varepsilon_{2,t})$.
\item Perform a Hadamard operation $\sigma^H_2$ at time $t=T=\tau+u$.
\end{compactenum}
This results in the two-qubit density matrix
\begin{eqnarray}
\rho_{2A}(u) &=& \Tr_S \sigma_2^H U_{S+A} [\chi^+,\chi^-]\sigma_2^H \rho_{S
+2A} \sigma^H_2 U^\dagger_{S+A}[\chi^+,\chi^-]  \sigma^H_2 \nonumber
\nonumber \\
&=& \left(1- \Re L_2(u)\,  \Sigma^z_2 - \Im L_2(u)\,  \Sigma^y_2 \right)/2
\label{eq:rho2A}
\end{eqnarray}
where 
\begin{eqnarray}
L_2(u) &=& e^{(i/\hbar) \int_0^{\tau+u}\varepsilon_{2,t} dt} G[u,\lambda] \label{eq:L2}\\
\Sigma^z_2 &=& \Pi_{-+}-\Pi_{--}\\
\Sigma^y_2 &=& |{-}{+} \rangle \langle {-}{-}|\,  +\,  |{-}{-} \rangle \langle {-}{+}| \, .
\end{eqnarray}
Thus performing two-qubit state tomography gives the characteristic
function $G[u,\lambda]$ of the process in Eq.~(\ref{eq:process}) at the
point $u$ apart from a known phase factor.
The state $\rho_{S+2A}$ can be prepared by thermalizing the $S+2A$ system at a 
temperature such that $\beta^{-1} \simeq \hbar \omega \ll \hbar \varepsilon_{1,0}, 
\hbar \varepsilon_{2,0}$. 

Two qubit-state tomography \cite{Liu05PRB72} can be realized in this setup by 
means of quantum non-demolition
joint dispersive read-out \cite{Filipp09PRL102b}. This is possible due to the fact 
that system and oscillator are far detuned.
Noticing that only terms involving $\Sigma_2^z$ and  $\Sigma_2^y$ appear in
Eq.~(\ref{eq:rho2A}), the wanted information can be retrieved in the
following way. (i) Follow the protocol describe above.  (ii) At the end of
the protocol, perform a measurement of the two-qubit observables
$\Sigma_2^z$ and  $\Sigma_2^y$.  Repeat (i) and (ii) many times to obtain
the expectation values $\langle \Sigma_2^z \rangle$, and $\langle
\Sigma_2^y \rangle$.  Then $\Re L_2(u) = - \langle \Sigma_2^z \rangle$,
$\Im L_2(u) =- \langle \Sigma_2^y \rangle$.

\subsection{Numerical solution}
We numerically studied the case of a linear ramp in the protocol
$\lambda_t= \lambda_0 + v t$ using $1/\omega$ as unit of time and the
parameters
$g_1=2.5  \omega$, $g_2=0.5  \omega$, $\lambda_0= 0.0625  \omega$,
$v= 1.5 \lambda_0 \omega/2 \pi$, $\tau = 2\pi/ \omega$. For $ \omega = 100$\,MHz this 
amounts to couplings 
$g_1 = 250$\,MHz and $g_2=50$\,MHz, an initial qubit splitting $\varepsilon_{1,0}
=10$ GHz 
and a velocity of $v \approx 150$\,(MHz)$^2$. The level splitting of the second qubit 
goes to infinity at the
beginning and at the end of the protocol, corresponding to a complete decoupling. 
The cutoff we introduced to handle this divergence is equivalent to $
\varepsilon_{2,0}=10$ GHz, which is within current experimental reach.
Fabrication of a slow oscillator with a slow frequency of $100$\,MHz does not seem 
to pose any special technological challenge. A slow mode oscillator can be made by 
increasing the resonator's length \cite{Duty05PRL}.
Perhaps more challenging is reaching the strong coupling $g_1=2.5 \omega$.
There is currently a strong interest in this regime of ultra-strong 
coupling, and we are optimistic that it will be soon reached
\cite{Peropadre10PRL105,Casanova10PRL105,Peropadre13arXiv}.
The time development of the two drivings $\chi^{\pm}_t$, is illustrated in Fig. 
\ref{fig:Fig3} bottom left. The graph in the bottom right panel of
Fig. \ref{fig:Fig3}, shows the corresponding time evolution of the two qubit energy 
splittings $\varepsilon_{i,t}$, $i=1,2$.
Note that $\varepsilon_{2,t}$ diverges for $t \rightarrow 0$ and for $t \rightarrow 
\tau+u$.
In our simulation, $\varepsilon_{2,t}$ was cut at the value of $100 \omega$. This 
results in a deviation of the actual drivings
$\chi^{\pm}_t$ from those reported in Fig.  \ref{fig:Fig3} bottom left panel, for those 
values of $t$ where the two $\chi$'s
approach. For small $u$ (as compared to $\tau$), this deviation becomes more 
relevant.
With the so chosen parameters, the condition (\ref{eq:regime2}) was obeyed at all 
times $t \in [0, \tau+u]$.

\begin{figure}[]
	\begin{center}
		\includegraphics[width=\textwidth]{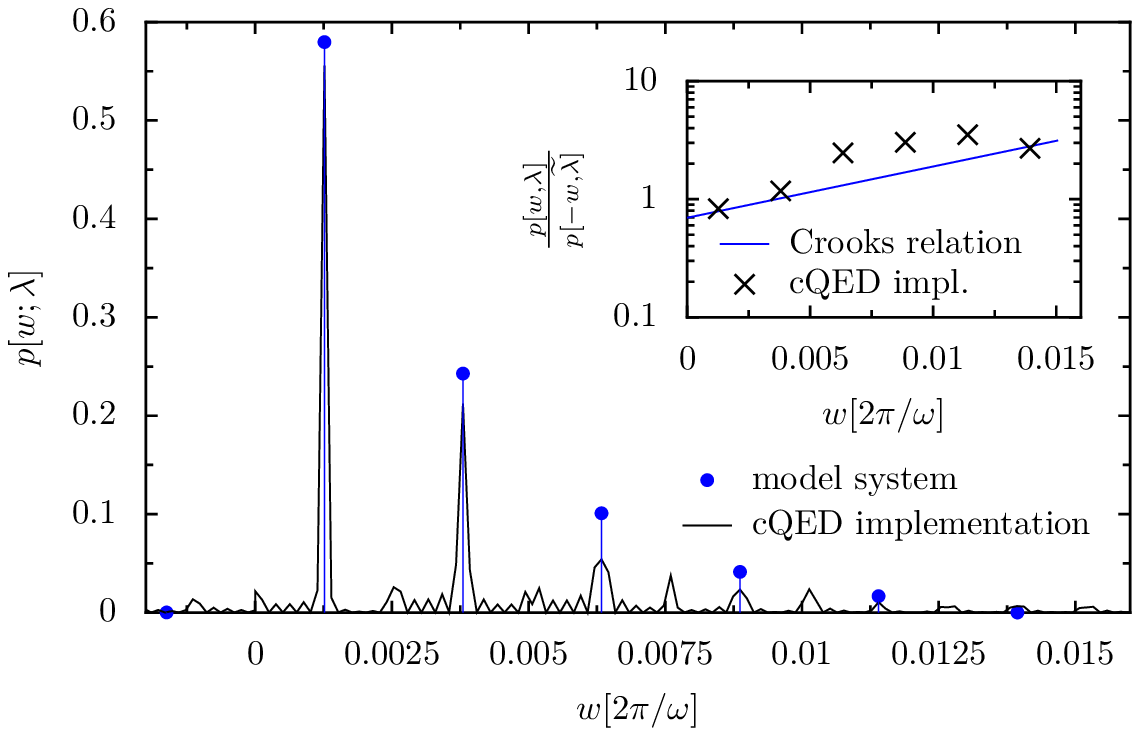}
		\caption{Work probability distribution of a parametrically driven 
oscillator. Solid black line:
		Numerical result obtained by the interferometric 2 qubits+oscillator 
setup. Blue points: Result obtained by solving
		the exact equations of motion governed by the Hamiltonian (\ref{eq:H-S}). 
		Inset: Check of the Crooks fluctuation theorem.
		The parameter used are: $g_1=2.5 \hbar \omega$, $g_2=0.5 \hbar 
\omega$, $v = 1.5 \lambda_0 \omega /2 \pi$, 
				$\lambda_0=0.0625 \hbar \omega$, $1/\beta= \hbar\omega$, 
$\tau = 2\pi/\omega$.}
		\label{fig:Fig4}
	\end{center}
\end{figure}

We computed $\rho_{2A}(u)$ according to Eq. (\ref{eq:rho2A}) where the time 
evolution was obtained by numerical integration of the 
Liouville-von Neumann equation. The thermal energy $\beta^{-1}$ was chosen 
equal to $\hbar \omega$.
We then extracted the real and imaginary parts of the characteristic function $G[u,
\lambda]$ using Eq.~(\ref{eq:L2}).
Figure \ref{fig:Fig4} shows the work probability distribution obtained after inverse 
Fourier transform of the so-obtained $G[u,\lambda]$.
The blue dots show the values of the work probability density function as obtained 
by integrating the model Hamiltonian (\ref{eq:process}) directly. The 
approximations introduced by our implementation result in a spread of the peaks, 
as compared to the expected ones, and to 
the emergence of further peaks in the work probability at high $w$ (not shown). 
Because of normalization these effects lower the height of the relevant peaks.
We repeated the same procedure for the time reversed protocol $\widetilde 
\lambda_t = \lambda_\tau - v t$. The inset of 
Fig. \ref{fig:Fig4} shows a good agreement between the logarithm of the
ratio $p[w;\lambda]/p[-w,\widetilde \lambda]$ as from our numerics, and the
linear behavior expected from Eq.~(\ref{eq:TC}).
The agreement is however not as good as one would expect from  Fig.
\ref{fig:Fig2} showing very good agreement between the dynamics of the
model Hamiltonian and actual Hamiltonian.
The source of error is coming from the fast oscillating phase $e^{(i/\hbar)
\int \varepsilon_{2,t} dt}$ in Eq. (\ref{eq:L2}), which has to be taken
away before the inverse Fourier transformation is applied.
This may pose an issue at the experimental level as well.

\section{Conclusions}
We have extended the interferometric scheme of
Dorner \emph{et al.} \cite{Dorner13PRL110} and Mazzola \emph{et al.}
\cite{Mazzola13PRL110} for the measurement of work distributions.
The method lends itself straightforwardly to the application
to open quantum systems, even in the regime of strong dissipation, which
represents a crucial advantage beyond the works by Dorner \emph{et al.}
\cite{Dorner13PRL110} and Mazzola \emph{et al.} \cite{Mazzola13PRL110}.  We
further showed how it can be modified to address the exclusive work
fluctuation theorem of Bochkov and Kuzovlev \cite{Bochkov77SPJETP45}.

Our central contribution is the illustration of a realistic
implementation of the method with current circuit QED technology. A new
feature of the proposed implementation is the introduction of a second
ancilla qubit and the use of two-qubit state tomography. This technique may
prove useful in all experimental scenarios where, as in the present case,
two independent drivings might not be easily achieved with a single qubit.
Our numerical calculations show the experimental feasibility.
In the proposed implementation, the driving
$\lambda_t (a^\dagger +a)^2$ is achieved indirectly by driving the qubit
splitting, and working in the soft mode regime (slow oscillator). As an
alternative to the proposed implementation one could control $\lambda_t$
directly. This can be implemented by coupling a flux qubit to a SQUID as
illustrated in Ref.~\cite{Peropadre10PRL105,Reuther11NJP9}.

\section*{Acknowledgments}
This work was supported by the German Excellence Initiative ``Nanosystems 
Initiative Munich (NIM)'' (MC, RB, PH), the Collaborative Research Center SFB 631 (RB, PH),
the Spanish Government grants MAT2011-24331 (SK) and 
FIS2011-25167 (DZ), the DGA project FENOL (DZ), the  EU project PROMISCE (DZ),
and the COST Action  MP1209 (MC, PH).

\appendix
\section{Derivation of the soft mode Hamiltonian $H'_{S+A}$ in Eq. 
(\ref{eq:rabi-g<<e-ad})}
The following derivation of the soft mode Hamiltonian is along the lines of
the derivation of the dispersive Hamiltonian presented by Schleich \cite{Schleich01Book}.
Our starting point is the Rabi Hamiltonian, Eq.~(\ref{eq:rabi}). For simplicity we do 
not include the time dependence of $\varepsilon$.
In the interaction picture, the qubit-oscillator coupling reads
\begin{equation}
H^I_{SA}(t) = \hbar g(\sigma_+ a e^{i\Delta t}+ \sigma_- a^\dagger
e^{-i\Delta t}+ \sigma_- a e^{-i\Gamma t}+ \sigma_+ ae^{i\Gamma t}) ,
\label{eq:HI-SA}
\end{equation}
where $\Delta = \varepsilon- \omega$, $\Gamma = \varepsilon + \omega$, and $
\sigma_\pm$ are the qubit rising and lowering operators. In the 
interaction picture, $H^I_{SA}(t)$ is the generator of the dynamics:
\begin{eqnarray}
U_{t,0}&=& T\exp \left( -\frac{i}{\hbar} \int_0^t dt' H^I_{SA}(t')\right)\nonumber \\
 &\simeq& 1 -\frac{i}{\hbar}  \int_0^t dt' H^I_{SA}(t') - \frac{1}{\hbar^2}
\int_0^t dt' H^I_{SA}(t') \int_0^{t'} dt'' H^I_{SA}(t'') .
 \label{eq:UI-SA}
\end{eqnarray} 
Plugging Eq.~(\ref{eq:HI-SA}) into (\ref{eq:UI-SA}), the first order term reads
\begin{eqnarray}
\hspace{-2.2cm}\int_0^t dt' H^I_{SA}(t') = \hbar g\left(
\sigma_+ a \frac{e^{i\Delta t} -1 }{i \Delta}   
+ \sigma_- a^\dagger \frac{e^{-i\Delta t} -1 }{-i \Delta} 
+ \sigma_- a \frac{e^{-i\Gamma t} -1 }{-i \Gamma} 
+ \sigma_+ a^\dagger \frac{e^{i\Gamma t} -1 }{i \Gamma} \right)  ,\nonumber \\
\label{eq:first-order}
 \end{eqnarray}
which can be used the to calculate the second order term
\begin{eqnarray}
\int_0^t dt' H^I_{SA}(t') \int_0^{t'} dt'' H^I_{SA}(t'') =  \hbar ^2g^2\int_0^t d t'   
\nonumber \\ \hspace{2cm}
 \left(  \sigma_+ \sigma_-  a a^\dagger \frac{1-e^{i\Delta t'} }{-i \Delta}
+ \sigma_+ \sigma_-  a^2 \frac{e^{i(\Delta-\Gamma) t'} -e^{-i\Delta t'}} {-i 
\Gamma}\right.  \nonumber \\ \hspace{2cm}
+ \sigma_- \sigma_+  a^\dagger a \frac{1-e^{-i\Delta t'} }{i \Delta}
+ \sigma_- \sigma_+ a^{\dagger 2}  \frac{e^{-i(\Delta-\Gamma) t'}-e^{-i\Delta 
t'} }{i \Gamma} \nonumber \\ \hspace{2cm}
+ \sigma_-\sigma_+  a^2 \frac{e^{i(\Delta-\Gamma) t'} -e^{-i\Gamma t'}} {i 
\Delta}
+ \sigma_- \sigma_+  a a^\dagger \frac{1-e^{-i\Gamma t'} }{i \Gamma}  
\nonumber \\  \hspace{2cm}
\left .+ \sigma_+  \sigma_- a^{\dagger 2} \frac{e^{-i(\Delta-\Gamma) t'} -e^{i
\Gamma t'}} {-i \Delta}
+ \sigma_+ \sigma_-  a^\dagger a\frac{1-e^{-i\Gamma t'} }{-i \Gamma} \right) .
 \end{eqnarray}
Recalling that the oscillator is slow compared to the qubit, for times short 
compared to the oscillator's period we can 
employ the approximation $e^{\pm i(\Delta -\Gamma)t'}= e^{-\pm 2i\omega t'}
\simeq 1$. In performing the integration 
we neglect the fast oscillating terms of frequency $\Gamma, \Delta \simeq 
\varepsilon$ to obtain
\begin{eqnarray}
\frac{1}{\hbar^2}\int_0^t dt' H^I_{SA}(t') \int_0^{t'} dt'' H^I_{SA}(t'')
\simeq \frac{i g^2}{\varepsilon} (a^\dagger + a)^2 \sigma^z t ,
\end{eqnarray}
where we used $\sigma_+ \sigma_- - \sigma_- \sigma_+ =\sigma^z$.
Note that the first order term, Eq. (\ref{eq:first-order}), contains either fast 
oscillating contributions or non relevant constant terms. Therefore it can be 
neglected at once, so that propagator becomes
\begin{eqnarray}
U_{t,0}  \simeq& 1 -\frac{i g^2}{\varepsilon} (a^\dagger + a)^2 \sigma^z t \simeq 
\exp\left(-i \frac{ g^2}{\varepsilon} (a^\dagger + a)^2 \sigma^z t \right)
\end{eqnarray}
which corresponds to the Hamiltonian $H^I_{S+A}(t) \simeq \frac{\hbar g^2}{\varepsilon} (a^\dagger + a)^2 
\sigma^z$. 
Going back to the Schr\"odingier picture, we finally arrive at $H_{S+A}\simeq \hbar 
\varepsilon \sigma^z/2 + \hbar (\omega a^\dagger a  +1/2) + \hbar g^2(a^\dagger + 
a)^2 \sigma^z/ \varepsilon $.
\section*{References}

\bibliographystyle{iopart-num}

\providecommand{\newblock}{}

\end{document}